# Water oxidation versus hydroxylation by the terminal oxo center of the $Fe^{III}$-hydroxide: DFT predictions

Aleksandr A. Shubin, Viktor Yu. Kovalskii, Sergey Ph. Ruzankin, Igor L. Zilberberg*, Valentin N. Parmon

*Boreskov Institute of Catalysis, Siberian Branch of the Russian Academy of Sciences, Prospekt Akademika Lavrentieva 5, Novosibirsk 630090, Russia*

## Abstract

The O-O coupling process is considered on terminal monocoordinated oxo centers in the gamma FeOOH hydroxide modeled by iron tetramer cubane cluster with the $Fe_4O_4$ core. The density functional theory predicts that reactive HO-$Fe^{IV}$-O• group formed from hydroxide by second withdrawal of proton-electron pair is capable to couple the OH moiety of water molecule with a low barrier. This process is far more effective than direct coupling of oxo centers on neighboring metal sites and is comparable with the coupling between terminal oxo center and three-coordinated lattice oxo center. The competing process of hydroxylation of oxyl oxygen to form the hydroxo group is equally probable having similar barrier.

## 1. Introduction

The hydroxides of transition metal are known to catalyze the water oxidation.[1] Nowadays investigations are focused mostly on extremely effective mixed (Ni,Fe) hydroxide for which the iron moiety is commonly considered as responsible for overall activity.[2][3] One of the major open questions in this field is the state of "active" iron cation and the detailed mechanism of the O-O coupling. The following experimental and DFT computations are known up to date.

On base of operando Mössbauer spectroscopic studies the $Fe^{IV}$ site of the (Ni,Fe) hydroxide was suggested to be responsible for the water oxidation [4]. The formation of such site within the ferryl $Fe^{IV}$=O species can appear via proton-coupled electron-transfer (PCET) from the $Fe^{III}$–OH species [5]. Alternatively, Goddard with coauthors suggested that the $Fe^{IV}$-O• species is a key intermediate determined activity of the (Ni,Fe) hydroxide.[6] Freibel, Bell, and Nørskov with coauthors on base of operando X-ray absorption spectroscopy (XAS) combined with high energy resolution fluorescence detection (HERFD) and DFT modeling came to conclusion that $Fe^{III}$ in $Ni_{1-x}Fe_xOOH$ is the actual active site for oxidation of water.[5] The ferric iron is claimed to occupy under-coordinated octahedral positions appeared on high-index surfaces $(01\bar{1}2)$ or $(01\bar{1}4)$ of NiOOH. [5]

* Corresponding author, E-mail: I.L.Zilberberg@catalysis.ru

As was suggested by Siegbahn, the O-O bond association on natural photosynthetic center necessarily involves an endergonic formation of oxygen radical [7]. Taking into account abovementioned ferryl configuration of active site, one may suggest that oxygen radical state appears on the way to transition states via the scheme shown for methane oxidation by ferryl oxygen [8][9] . Nevertheless, the detailed mechanism of the O-O coupling is still unknown.

In our previous work, with the use of tetramer model $Fe_4O_4(OH)_4$, assuming that the first PCET forms the ferryl $Fe^{IV}$=O moiety from $Fe^{III}$-OH group, the most energetically favorable route for the O-O bond formation was shown to be that associated with the incorporation of ferryl oxygen to the tetramer edge with a barrier of 12 kcal/mol [8]. A competing process blocking this scenario is the water hydroxylation of the ferryl center to form two HO-$Fe^{IV}$-OH instead of ferryl group. However, second PCET from this group to form HO-$Fe^{IV}$-O• or HO-$Fe^{V}$=O groups again "opens" the terminal oxo center capable of the O-O coupling with lattice oxo center. In addition, the presence of hydroxo group in the neighborhood of terminal oxo creates a possibility of the oxo-hydroxo association at the same Fe center to form the OOH species. So-obtained O-O coupling on a single iron site is though less probable due to a relatively high barrier of 18 kcal/mol [8].

As far as the above described formation of the -O• or =O terminal oxo centers is concerned, a question arises whether unavoidable hydroxylation process can deactivate these oxo sites in water solution. One may guess that electrophilic attack of water molecule on oxo sites resulting in the $Fe^{IV}$=OH and $Fe^{V}$=OH formation competes with the nucleophilic water attack on the same sites to form hydroperoxo species Fe-OOH. To answer this question the DFT comparative modeling of the hydroxylation and oxidation has been performed using simple cubane cluster $Fe_4O_4(OH)_4$ used in our previous works.

## Model

Active sites of water-oxidation catalysts based on iron hydroxides are believed to have much in common with the structure of gamma-FeOOH hydroxide. The latter consists of the double chain of edge-sharing $Fe(O,OH)_6$ octahedra. Major structural motif of this Fe-hydroxide is a trimer consisting of iron-centered octahedra (Figure 1). The Fe-O-Fe angle is about 100 degrees. Oxo centers are always three-coordinated, while hydroxo groups couple two or one Fe cation. Monocoordinated hydroxyl can appear only on the vertice of the terminating octahedron and is most probably the subject of first PCET step initiating various scenarios for the O-O coupling process.

The cubane tetramer (in terms of Fe atoms) has been used for modeling. Such model was proved to be quite useful in modeling O-O coupling allowing to simulate oxidation acts and O-O coupling on the vertex of terminal octahedron in the edge-sharing $M(O,OH)_6$ (M=Co,Fe) octahedra chain. [10][11][8]

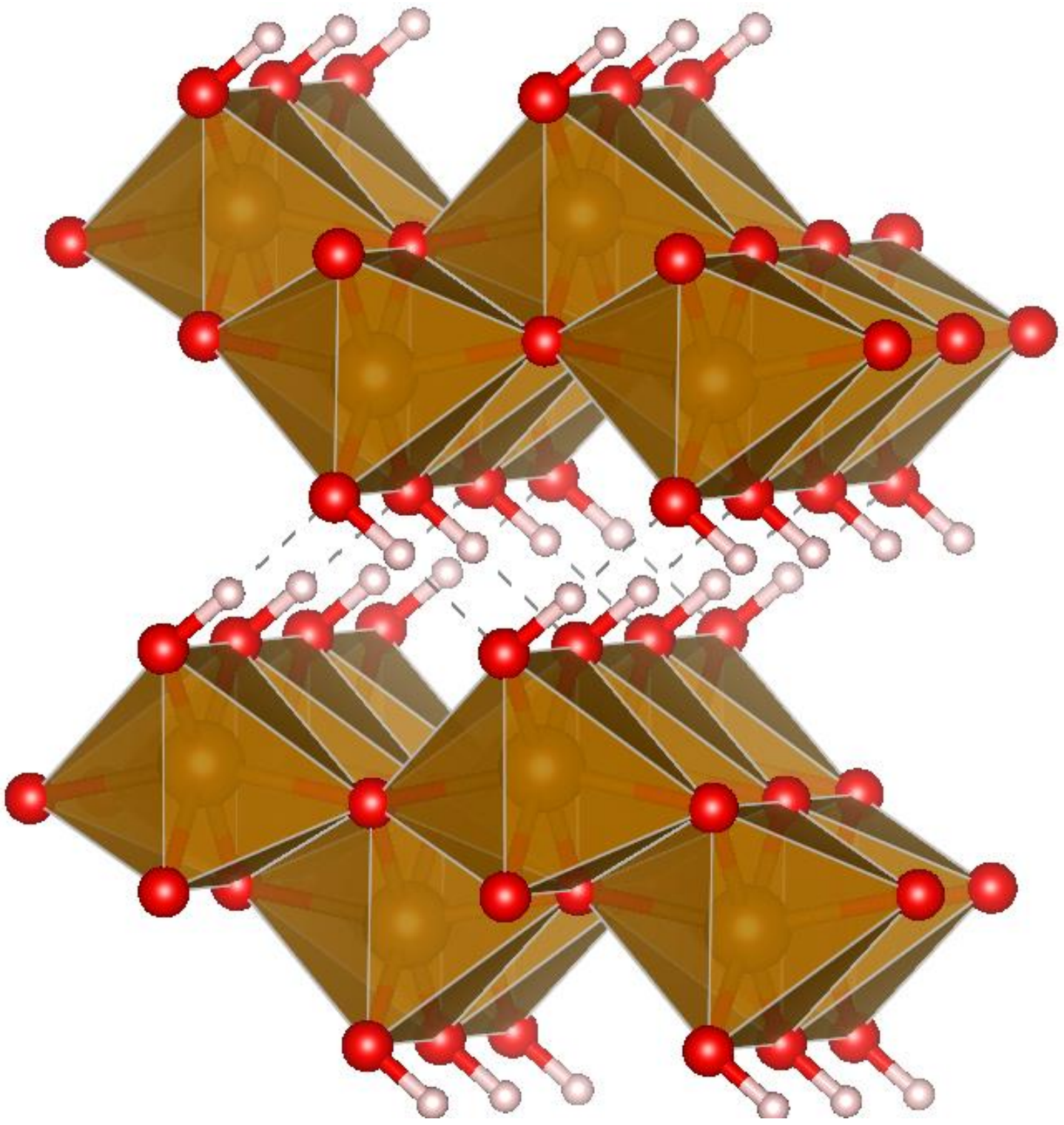

**Figure 1.** γ-FeOOH hydroxide structure: iron centers are hidden in octahedra, oxygen centers are in red, hydrogen centers are in white

The abstraction of proton and electron from vertex monocoordinated hydroxyl group forms the ferryl group $Fe^{IV}$=O which can then relatively easily (with a barrier of 12 kcal/mol) couple three coordinated lattice oxo center to form peroxide species. Further release of molecular oxygen from this peroxide species faces no any substantial barriers. [8] However, far more active terminal oxo center can appear at the second step of proton-electron pair withdrawal. As was shown in our work, the ferryl group can be a target of nucleophilic water molecule attack resulting in the appearance of two hydroxyls instead of ferryl group:

$H_2O + Fe^{IV}{=}O \rightarrow HO\text{-}Fe^{IV}\text{-}OH$

The removal of electron-proton pair from one of these hydroxyls forms another bare terminal oxygen (of the oxyl or ferryl type) on the same iron cation:

$HO\text{-}Fe^{IV}\text{-}OH – (H^+,e) \rightarrow HO\text{-}Fe^{IV}\text{-}O\bullet$ or $HO\text{-}Fe^{V}{=}O$.

So-formed group in oxyl configuration $HO\text{-}Fe^{IV}\text{-}O\bullet$ can be apparently quite reactive toward the O-O coupling process due to high value of β-spin density on it.

With these reasons in mind, the model reactive center was chosen to be above mentioned cubane cluster having terminal oxo and hydroxo ligands at the corner (Figure 2). The total spin projection of this cluster is set to 9 on base of the following data. For previously considered tetramer $Fe_4O_4(OH)_4$ the lowest total energy corresponds to maximal spin of 10.[8] Removal of first proton-electron pair from hydroxyl group to form terminal ferryl oxo center decreases spin to 19/2. Hydroxilation of ferryl center followed by second proton-electron pair removal further decreases spin to 9. Formal scheme of oxidation and spin states for obtained tetramer core is most likely $Fe^{V/IV}$(S=3/2) $Fe^{III}$(S=5/2)$Fe^{III}$(S=5/2)$Fe^{III}$(S=5/2) with the reactive iron being in competing ferryl $Fe^{V}$ and oxyl $Fe^{IV}$ states corresponding to spin up and spin down on terminal oxo center, respectively.

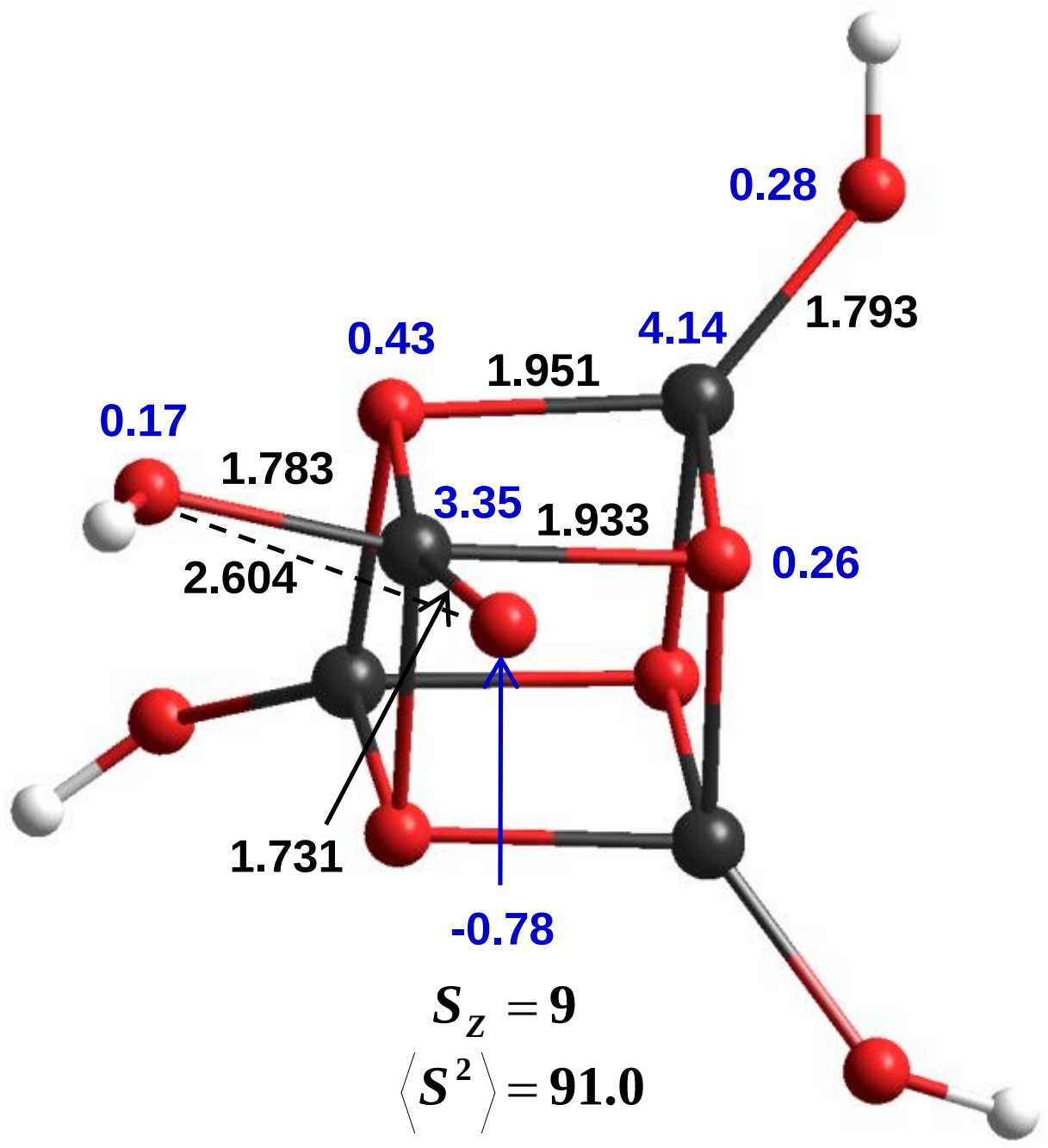

**Figure 2.** Gase-phase cubane model of the HO-Fe-O corner (iron in black, oxygen in red, hydrogen in white). The listed values with three and two decimal digits show the distances in Å (in black), and Mulliken spin density (in blue) on atoms, respectively

## 2. Computational details

All calculations have been performed at the UB3LYP/6-311G++(d,p) level with ultrafine integration grid within the framework of the Gaussian09 package [12]. For starting complex $Fe_4O_4(OH)_4$ the solution for the spin projection of 10 (which means parallel spins on iron centers) appears to have minimal energy among all possible iron spins configurations. Ferromagnetic coupling of spins on metal centers seems to be a sequence of cubic geometry of tetramer $Fe_4O_4(OH)_4$ having right angles Fe-O-Fe in a perfect agreement with the prediction on negligible superexchange for the $Fe^{3+}$-$O^{2-}$ -$Fe^{3+}$ linkage at the 90° angle.[13,14] For important processes the presence of water solvent was accounted within Polarizable Continuum Model (PCM) using the integral equation formalism variant (IEFPCM) [15] which is the default SCRF method in Gaussian09 package.

## 3. Results and discussion

### 1. Hydroxylation

When terminal oxyl oxygen in the $Fe^{IV}$-O• group abstracts hydrogen from water molecule adsorbed on neighboring metal center (Figure 3), resulting structure appears to contain two hydroxyl groups on metal sites (Figure 3), rather than the Fe-OH moiety and free OH radical as it might be expected for the water oxidation route. Therefore, this process (having a low barrier of 9 kcal/mol) has to be assigned to the dissociative adsorption or hydroxylation. The same is true for nucleophilic water attack on ferryl oxygen (Figure 4). For the latter case, the abstraction of hydrogen on ferryl oxygen to form two hydroxyl anions goes through even a lower barrier of 4 kcal/mol (Figure 4). This might be explained by more nucleophilic nature of ferryl oxygen preferable for abstracting proton, the fact which is seen from the difference between the energy of 1s(O) level for ferryl and oxyl oxygen. The latter is 1.2 eV lower than the former implying less negatively charged oxyl oxygen (Table S1)

Although the hydroxilation of oxyl oxygen in the HO-$Fe^{IV}$-O• group and neighboring $Fe^{III}$ center forms two chemically equivalent centers having two hydroxo ligands, their spin density (and so oxidation state) remains almost the same (Figure 3). However, the change takes place for the iron center which is directly not involved in the process. Its spin density drops from 4.21 to 3.40. Taking into account that energies of the 1s(Fe) level for the iron centers with spins 3.34 and 3.40 are equal within 0.01 eV (Figure 3c), one might guess that the oxidation state of the “spin 3.40” iron is $Fe^{IV}$ as in case of the iron with oxo ligand (Figure 3a). What is interesting, initial oxyl containing cubane has oxidation states $Fe_4$(IV,III,III,III) (Figure 3a). Therefore, hydroxilation changes this configuration to $Fe_4$(IV,IV,III,III) (Figure 3c) implying delocalization of spin over the oxo centers. This result reveals unusual effect that the dissociative adsorption of water (which normally proceeds without any electron transfer) on cluster affects oxidation states of connected iron centers. Certainly, this effect is connected with the partial disruption of cubane structure as seen from the elongation of one of edges by almost 1 Å (Figure 3c). In case of hydroxylation of the ferryl-oxo cubane the oxidation scheme is $Fe_4$(IV,IV,III,III) from the beginning at each steps of the process (Figure 4).

Worthwhile noting that account of solvation does not change much the structure and relative energies of above given process of water dissociation. For the processes (in both oxyl and ferryl cases) modeled without such account, the barrier is only 1 kcal/mol larger than that for the model with solvation (Figure S2).

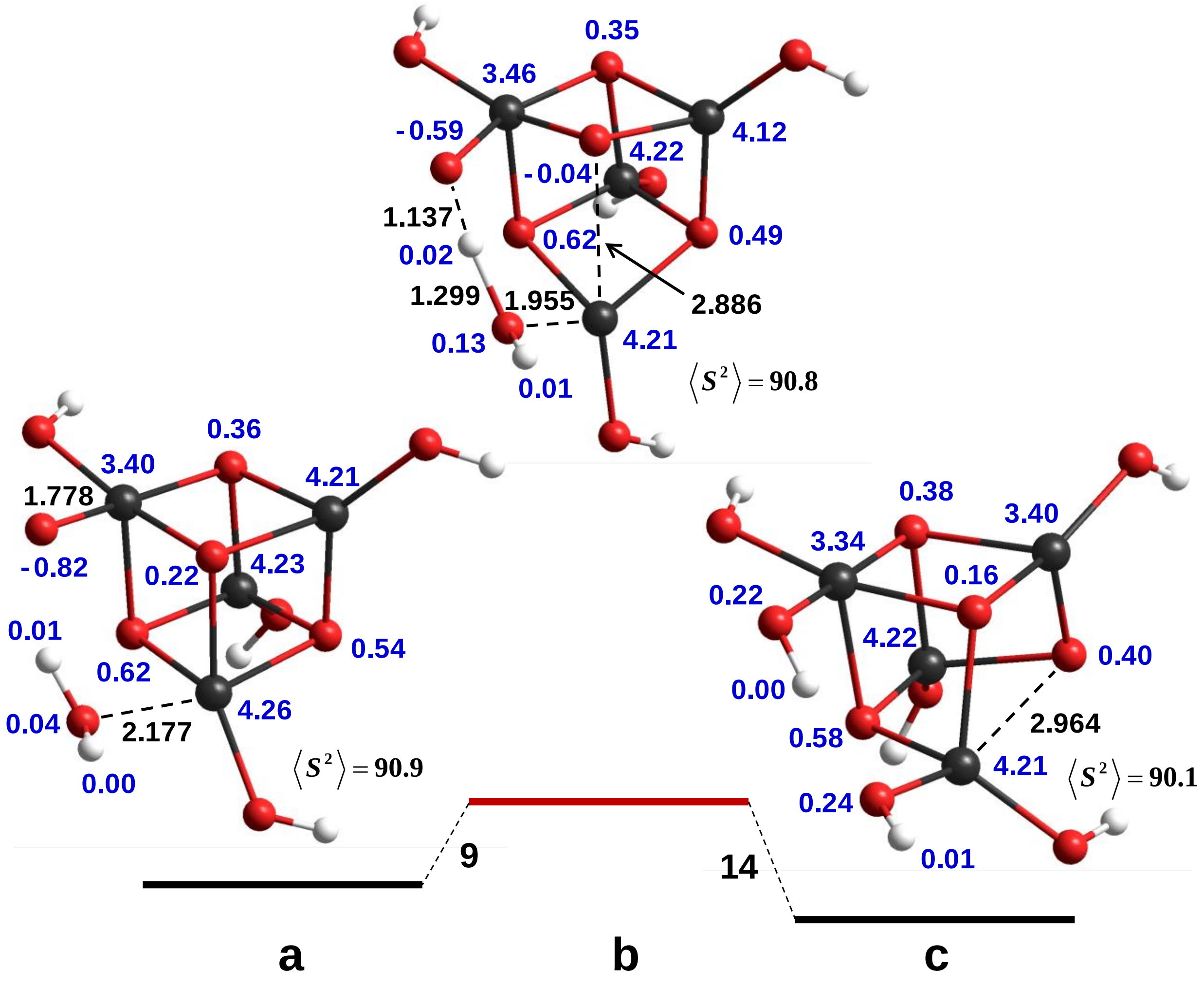

**Figure 3.** Dissociative chemisorption of water molecule on oxyl oxygen center: starting complex, transition-state complex and resulting hydroxylated complex. Integer numbers in energy diagram list the total energy differences (in kcal/mol).

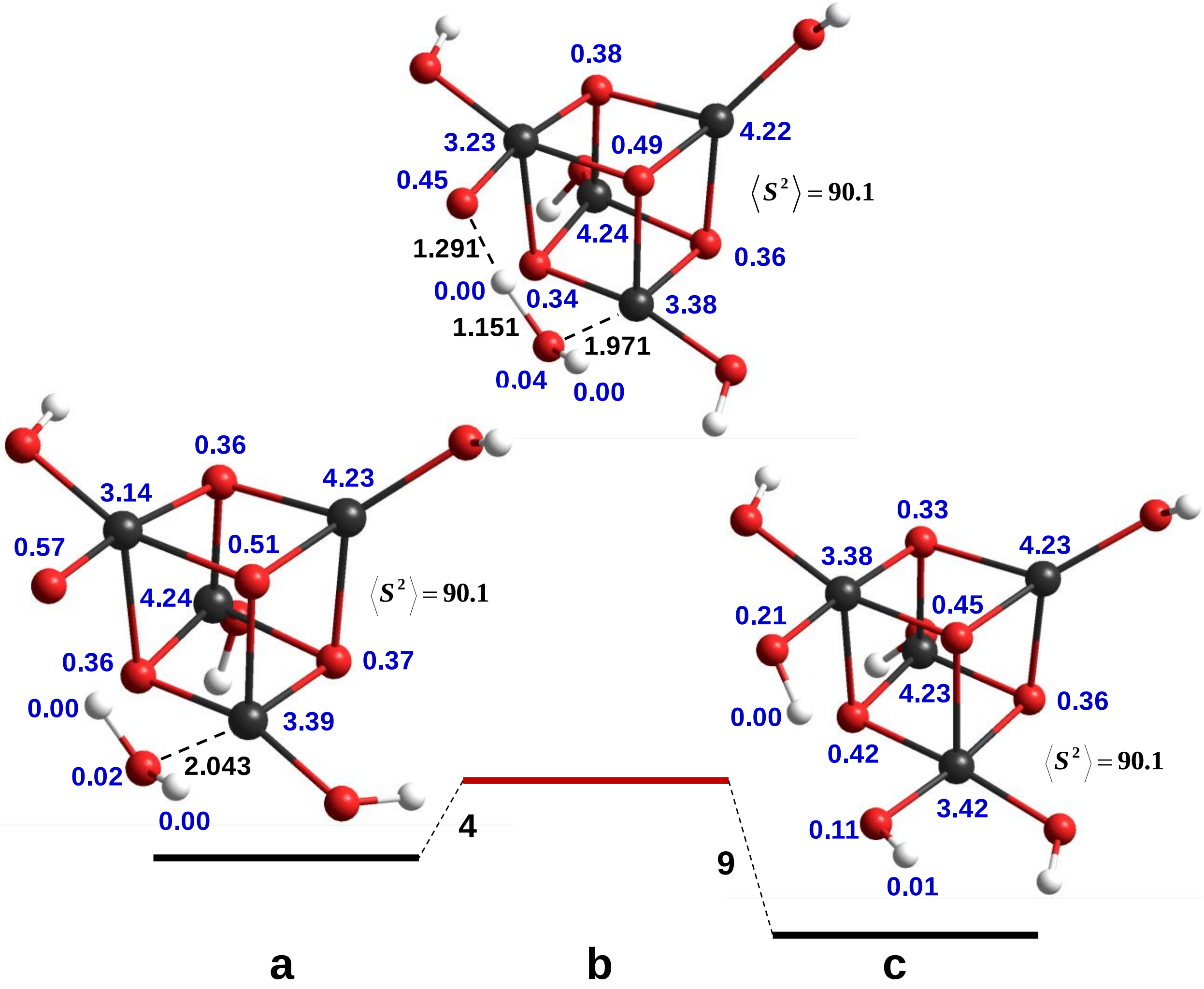

**Figure 4**. Dissociative chemisorption of water on ferryl center ferryl oxygen in $Fe^{IV}$=O.

## 2. Oxidation of water

True O-O coupling takes place for the water nucleophilic attack on the terminal oxyl oxygen center (Figure 5). The process starts from the formation of hydrogen bonds between water molecule and the OH ligand of reactive iron center (Figure 5a). Two simultaneous steps takes place: penetrating of hydrogen atom from water to hydroxyl ligand and associating of remaining OH group to terminal oxyl oxygen. The barrier of 11 kcal/mol for coupling of water oxo center to oxyl center looks amazingly low in comparison with barriers of 22-43 kcal/mol for direct coupling of oxo centers on neighboring iron centers as we found in our previous work (Figure 5-7 in ref. [8]). Especially interesting the comparison with the OOH group formation without water with barrier18 kcal/mol (Figure 4 in ref.[8]). Present work reveals an effect of upcoming water molecule which catalyzes in fact this process dropping the barrier by 7 kcal/mol.

Initial scheme of oxidation states for cluster with oxyl oxygen is formally $Fe_4$(IV,III,III,III,III). It becomes $Fe_4$(II,III,III,III,III) as seen from dropping of spin density on reactive iron from 3.40 to 2.83 (Figure 5ac) and corresponding rising of the core energy $\varepsilon_{1s}$(Fe) by 2.2 eV implying substantial "back" transfer of electron density from oxo and hydroxo ligands into iron center.

Above described O-O coupling is obtained for five coordinated reactive iron center one might suspect that a low barrier is an artifact. To clarify this issue additional molecule was put to form six-coordinated iron center of cubane (Figure S8a) and all the process has been modeled again. The barrier of the O-O coupling appears to be 9 kcal/mol, in fact coinciding with the results for coordinatively unsaturated iron center. This is not surprising as the actual hydrogen transfer between water and hydroxo group and simultaneous coupling OH takes place in a close proximity of oxo and hydroxo groups. Moreover, the barrier seems to be determined by the ability of reactive iron center and its immediate neighbors to adopt formally two electrons from terminal oxo center and hydroxo groups becoming OOH and water ligands. Quite evident that water solvent could not bring noticeable contribution into this process.

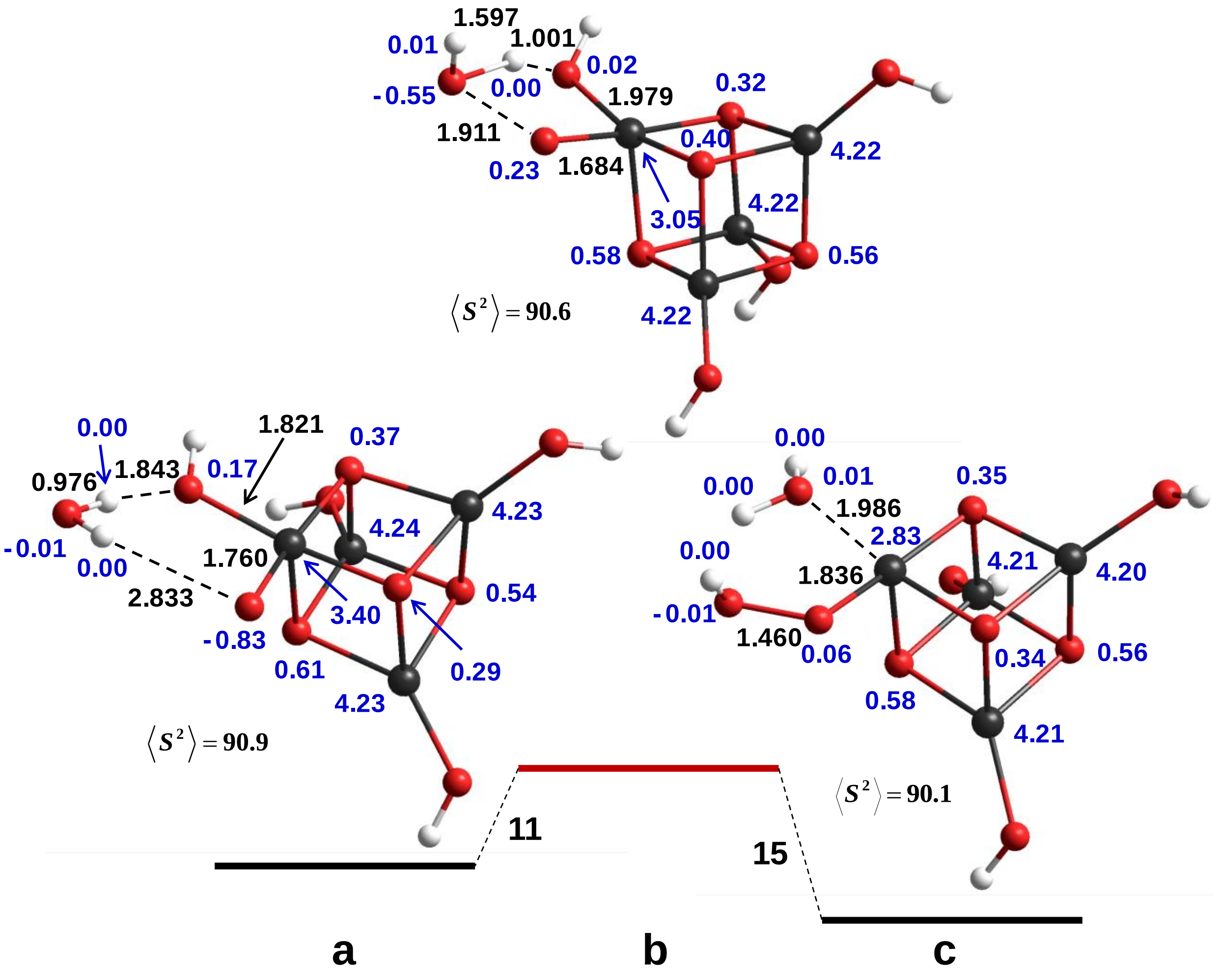

**Figure 5.** Nucleophilic attack of water molecule on oxyl oxygen in FeIII-O• to form Fe-OOH.

## 4. Conclusion

In the present work the cubane cluster $OFe_4(\mu\text{-}O)_4(OH)_5$ was used to model the O-O coupling versus hydroxylation of the reactive terminal oxo center on the iron-containing oxyhydroxides by means of DFT with solvent account. The following results are obtained.

The water molecule attacking reactive corner site hydroxylates both oxo centers (HO-$Fe^{IV}$-O• and HO-$Fe^{V}$=O) to form hydroxo group from terminal oxo group as well as hydroxo group on neighboring iron site:

$H_2O$ + •O(OH)$Fe^{IV}$-O-$Fe^{III}$(OH) → $(HO)_2Fe^{IV}$-O-$Fe^{III}(OH)_2$

$H_2O$ + O(OH)$Fe^{V}$-O-$Fe^{IV}$(OH) → $(HO)_2Fe^{V}$-O-$Fe^{IV}(OH)_2$.

The activation barrier for hydroxylation of oxyl oxygen is predicted to be 9 kcal/mol, while in case of ferryl oxygen the barrier is 4 kcal/mol.

The nucleophilic water attack on the oxyl oxygen to form OOH group proceeds with a comparable barrier of 11 kcal/mol:

$H_2O$ + •O(OH)$Fe^{IV}$-O-$Fe^{III}$(OH) → HOO-$Fe^{II}$-O-$Fe^{III}(OH)_2$.

Obtained barrier estimations allows one to conclude that hydroxylation and O-O coupling are equally probable. From two forms of terminal oxo center the ferryl one is preferred for hydroxylation. Interestingly, that this center is less preferred for O-O coupling. Therefore, the hydroxylation in fact enhances selectivity of the O-O coupling on the oxyl-oxygen centers.

## Acknowledgement

Calculations have been performed at the Siberian Supercomputer Centre SB RAS. This research was supported by Russian Foundation for Basic Research under grant No. 15-29-01275.